\documentclass{article}

\usepackage{microtype}
\usepackage{graphicx}
\usepackage{subfigure}
\usepackage{booktabs} 

\usepackage{siunitx}
\DeclareSIUnit\knot{kn}

\usepackage{hyperref}

\usepackage[accepted]{icml2024}

\usepackage{amsmath}
\usepackage{amssymb}
\usepackage{mathtools}
\usepackage{amsthm}

\usepackage[capitalize,noabbrev]{cleveref}

\theoremstyle{plain}

\theoremstyle{definition}

\theoremstyle{remark}

\usepackage[textsize=tiny]{todonotes}

\icmltitlerunning{Satellite to Radar Vision Transformer (SRViT)}

\begin{document}

\twocolumn[
\icmltitle{SRViT: Vision Transformers for Estimating\\Radar Reflectivity from Satellite Observations at Scale}



\icmlsetsymbol{equal}{*}

\begin{icmlauthorlist}
\icmlauthor{Jason Stock}{csu}
\icmlauthor{Kyle Hilburn}{cira}
\icmlauthor{Imme Ebert-Uphoff}{cira,ece}
\icmlauthor{Charles Anderson}{csu}
\end{icmlauthorlist}

\icmlaffiliation{csu}{Computer Science, Colorado State University, USA}
\icmlaffiliation{cira}{Cooperative Institute for Research in the Atmosphere, CO, USA}
\icmlaffiliation{ece}{Electrical and Computer Engineering, Colorado State University, USA}

\icmlcorrespondingauthor{Jason Stock}{stock@colostate.edu}

\icmlkeywords{Machine Learning, Transformers, Deep Learning, Weather and Climate, Atmospheric Science}
\vskip 0.3in
]



\printAffiliationsAndNotice{}  

\begin{abstract}
We introduce a transformer-based neural network to generate high-resolution ($3$ km) synthetic radar reflectivity fields at scale from geostationary satellite imagery. This work aims to enhance short-term convective-scale forecasts of high-impact weather events and aid in data assimilation for numerical weather prediction over the United States. Compared to convolutional approaches, which have limited receptive fields, our results show improved sharpness and higher accuracy across various composite reflectivity thresholds. Additional case studies over specific atmospheric phenomena support our quantitative findings, while a novel attribution method is introduced to guide domain experts in understanding model outputs.
\end{abstract}

\section{Introduction}

Accurate radar is crucial for operational forecasters to monitor and forecast the progression of high-impact weather events. Given the implications to public safety and agriculture, among others, this accuracy is key to protecting life and property. There are two primary uses of radar in weather forecasting: (1) enabling forecasters to view imagery and decide when to issue warnings and (2) for imagery to be integrated into numerical weather prediction (NWP) models to forecast the weather \cite{gustafsson2018survey,jones2020assimilation}. However, radar is limited to sparsely situated ground stations, leaving remote areas, mountains, and oceans with poor coverage. (\cref{fig:spatial-distribution}; see Figure 1 in \citet{mcgovern2022we}). To achieve more accurate forecasts, it is essential to improve radar coverage both spatially and temporally, which we aim to address by using observational data.

\begin{figure}[t!]
    \centering
    \includegraphics[width=1\linewidth]{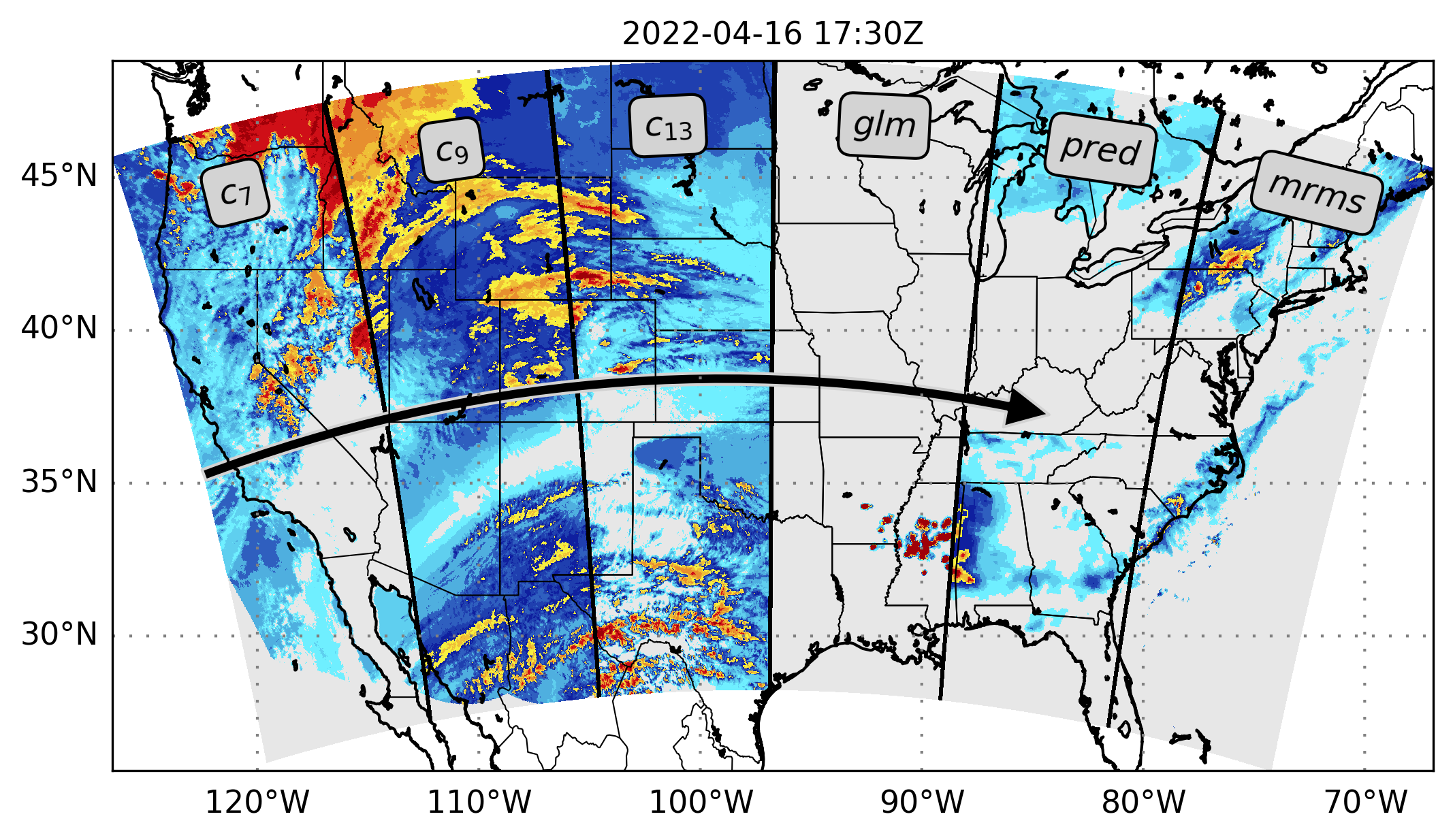}
    \vspace{-6mm}
    \caption{Slices of an input (col. 1-4) and output prediction with ground truth (col. 5 and 6, respectively). The real-time observations enable forecasters to assess/forecast storm patterns at scale.}
    \label{fig:output}
\end{figure}

\begin{figure*}[t!]
    \centering
    \includegraphics[width=1\textwidth]{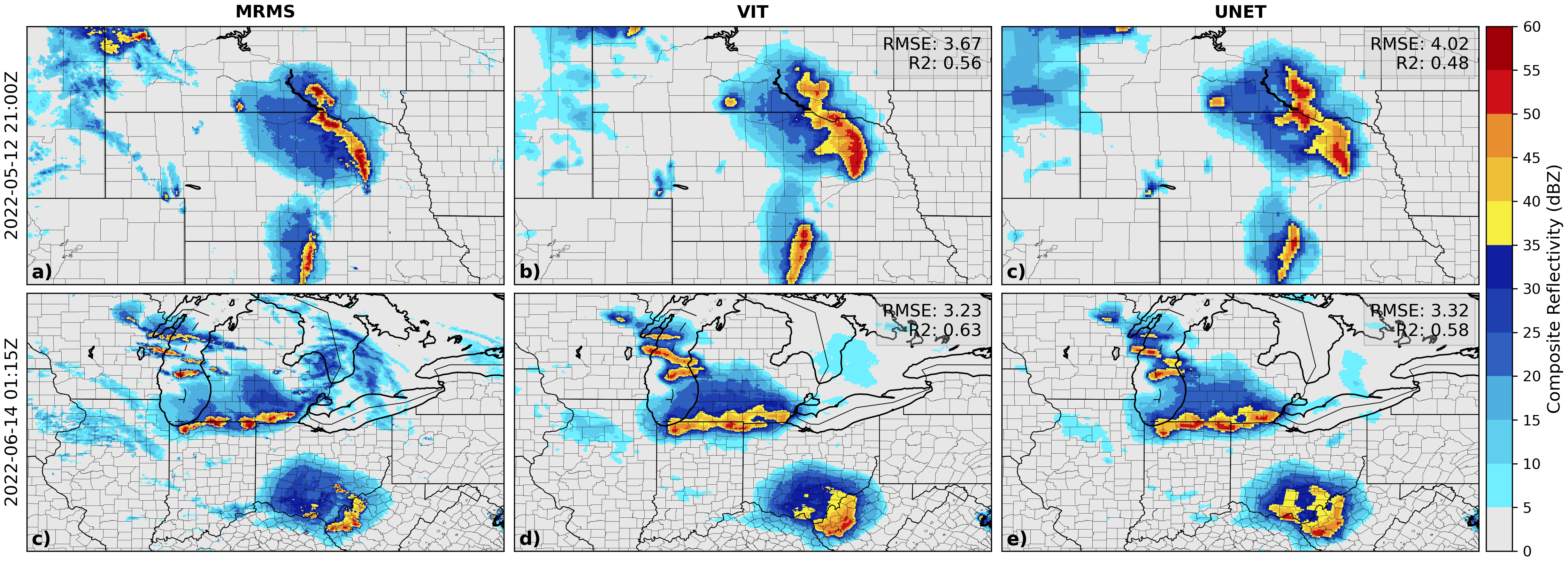} 
    \vspace{-4mm}
    \caption{Cropped and enlarged model output, showing a Northern Plains Derecho in panels (a-c) and Midwest Squall Lines in panels (d-f). Sample RMSE and $\text{R}^2$ values are shown for each case between the ground truth MRMS (col. 1) and model output (col. 2 and 3).}
    \label{fig:cases}
\end{figure*}

The Geostationary Operational Environmental Satellite (GOES) provides expansive coverage of the contiguous United States (CONUS) with $5$ minute updates and has been shown to be effective for operational weather forecasting \cite{line2016use}. With its low-latency and high spatial coverage, we believe there is a great opportunity in leveraging its geostationary imagers. In particular, our study focuses on utilizing imagery from the Advanced Baseline Imager (ABI) \cite{schmit2010goes} and the Geostationary Lightning Mapper (GLM) \cite{goodman2010glm} onboard GOES-16 to generate high-resolution synthetic radar as modeled by the Multi-Radar Multi-Sensory (MRMS) product \cite{smith2016multi} over the CONUS domain (\cref{fig:output}). 

We use a machine learning approach, a field that has recently received attention for radar estimation. However, previous work has centered on incorporating physical indicators from numerical models \cite{veillette2018creating,zhu2023multiscale} or modeling localized spatial regions \cite{Hilburn2021Gr,duan2021reconstruction,yu2023radar}. Despite the influx of high-resolution weather and climate modeling \cite{pathak2022fourcastnet,mardani2024residual}, the transition to high-resolution observational-based modeling is still in its infancy \cite{stock2024diffobs}. In this work, we hypothesize that transformer-based models can benefit from contextualizing the synoptic observations over what is possible with convolutional models. We not only evaluate our results both quantitatively and through case studies, but we also introduce an attribution method for transformers to better assist domain experts.

\section{Preliminaries}

We begin by introducing the dataset in \cref{sec:composition}, detailing our data features and composition, and in \cref{sec:baseline} we outline the baseline model used for making comparisons.

\subsection{Data Composition} \label{sec:composition}

This study uses observational data as \textit{input} (ABI infrared Channels 7, 9, 13 and GLM) from the Geostationary Operational Environmental Satellite (GOES)-R Series satellites with \textit{target} composite radar reflectivity from the Multi-Radar Multi-Sensor (MRMS) product---complete variable details are provided in \cref{app:dataset.varaibles}. We collect data in part from \citet{2020conus3,2021conus3,2022conus3} between 2018-2022 that span the contiguous United States (CONUS). As the intent of our results are to be used with data assimilation, we project the input and target samples to follow a $3$ km HRRR mass grid, yielding $768 \times 1536$-pixel images ($2304 \times 4608$ km) on $6$ hour periods with a $15$ minute refresh ($96$ samples per day). 

We restrict data to the warm season (i.e., April-September) where the strongest radar echos occur. Data are temporally partitioned to the years of 2018-2020 ($47{,}821$) for training, 2021 ($17{,}284$) for validation, and 2022 ($17{,}344$) for testing. Note the number of samples are indicated in parentheses that altogether total $1.9$ TBs in size. All samples undergo a preliminary quality control phase to ensure targets have the full range of radar reflectivity and there are no erroneous samples with large GLM and MRMS artifacts (additional verification in \cref{app:dataset.coverage}). Additionally, individual variables are scaled between $[0,1]$ based on their corresponding histograms to stabilize training. When unstandardized, MRMS and model output values scale between $[0,60]$ dBZ (unit of reflectivity).

\subsection{Baseline Model Comparison} \label{sec:baseline}

The GOES Radar Estimation via Machine Learning to Inform NWP (GREMLIN) model \cite{Hilburn2021Gr} uses a fully-convolutional encoder/decoder architecture (without skip connections) that we denote as ``UNet'' in this study. Trained previously on smaller $256 \times 256$-pixel images and regionally filtered by storm reports (tornado, hail, and wind events), this architecture serves as the baseline for satellite to radar emulation. Using the same hyperparameters as in the original work, we retrain the UNet and both evaluate and compare its performance to our model using data from the entire CONUS domain.

\begin{figure*}[t!]
    \centering
    \begin{minipage}[b]{0.475\textwidth}
        \centering
        \includegraphics[width=0.9\columnwidth]{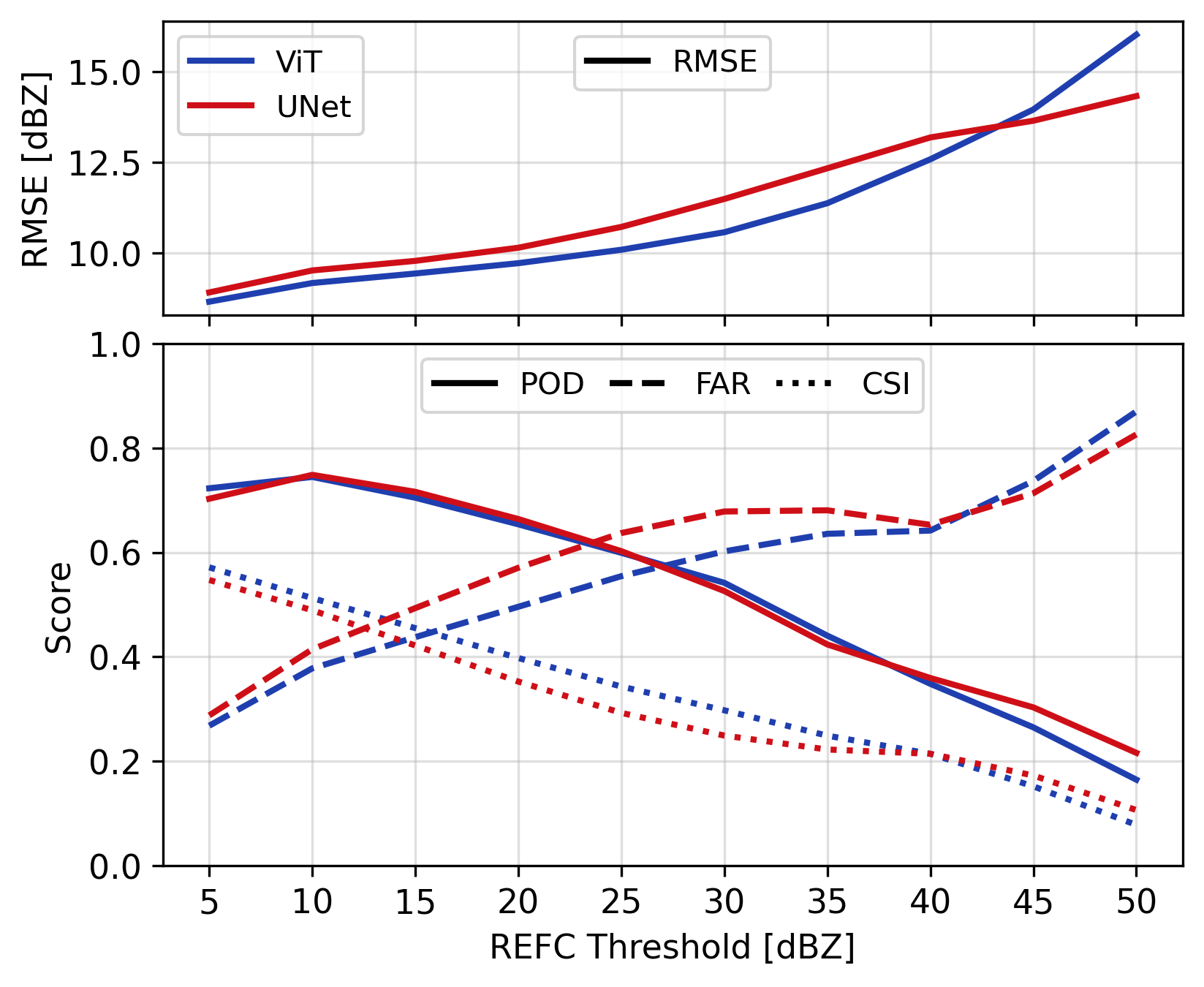}
        \caption{Categorical metrics at varying composite reflectivity thresholds for SRViT and the baseline UNet.}
        \label{fig:stats}
    \end{minipage}
    \hfill
    \begin{minipage}[b]{0.475\textwidth}
        \centering
        \includegraphics[width=0.9\columnwidth]{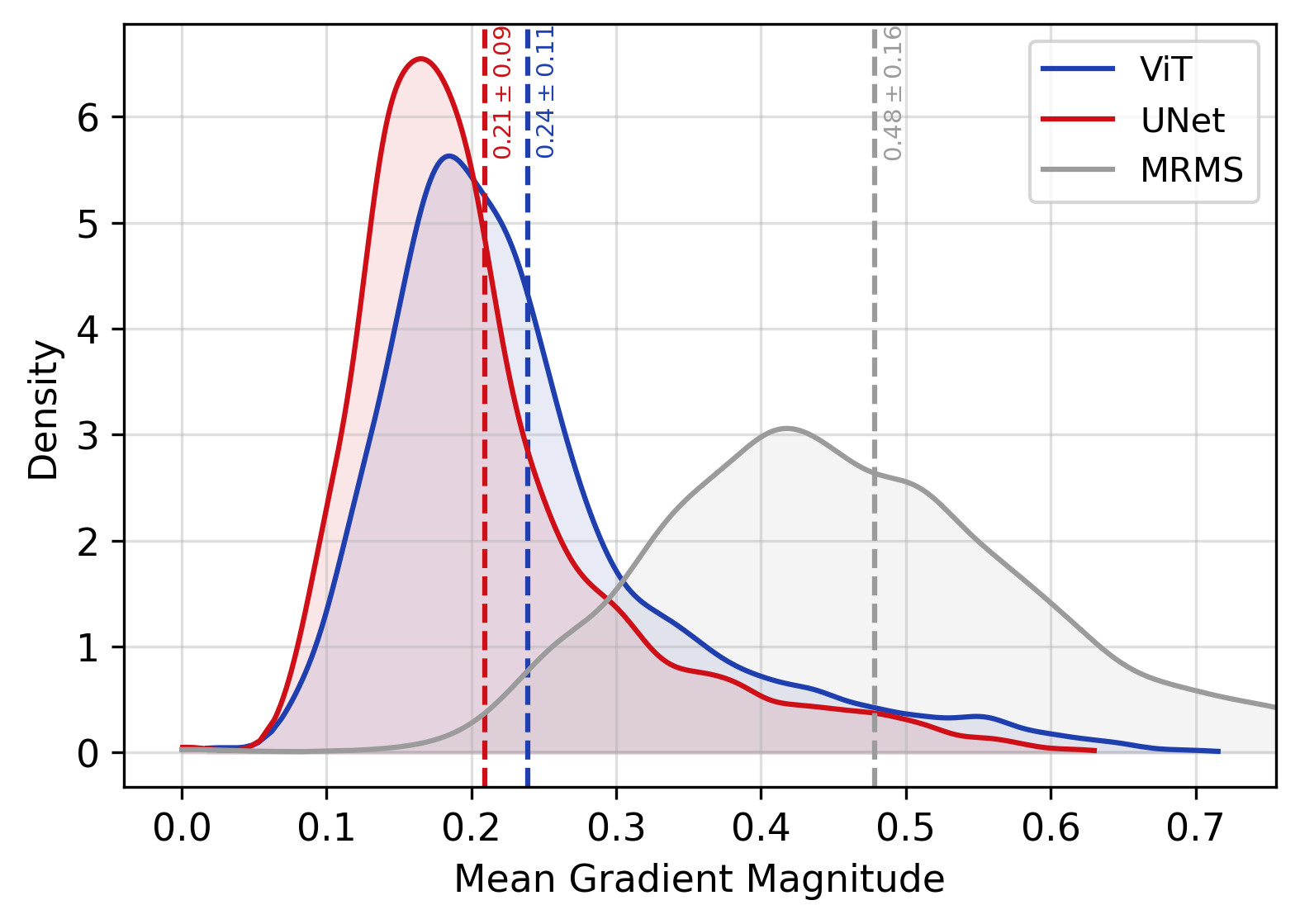}
        \caption{Kernel density estimation (KDE) of the mean gradient magnitude of composite reflectivity over all \textit{test} samples. The dashed line represents the mean with standard deviation.}
        \label{fig:grad}
    \end{minipage}
\end{figure*}

\section{Methodology} \label{sec:methodology}

Our proposed model extends the vision transformer \cite{vaswani2017attention,dosovitskiy2020image} and is designed specifically for image-to-image translation. This is motivated by wanting to address the limitations of restricted receptive fields and empirical evidence for blurriness in tandem with pixel-wise loss functions \cite{ravuri2021skilful,ayzel2020rainnet} with purely convolutional-based methods. By using self-attention, we are able to contextualize the long-range dependencies and synoptic information of global weather patterns that influence local precipitation events. 

The design specifics of our Satellite to Radar Vision Transformer (SRViT), as detailed in \cref{app:model}, is inspired by prior global weather forecasting models. In particular, work from \citet{pathak2022fourcastnet,bi2022pangu,chen2023fengwu} that all use transformer-based architectures and are shown to outperform graph- and convolutional-based approaches while approaching the performance of NWP \cite{lam2022graphcast,rasp2021data}. However, in this work we simplify the architecture and training procedure.

We use the weighted loss from \citet{Hilburn2021Gr} to balance the rare but high radar reflectivity values with the small but common values. Specifically, denoted as
\begin{align} \label{eq:loss}
    \mathcal{L}_{e} = \frac{1}{m}\sum_{i=1}^m {\exp{(w_0 t_i^{w_1})}\cdot(y_i - t_i)^2},
\end{align}
where $t$ and $y$ are the ground truth and predicted values, $m$ is the number of pixels, and $w_0=5$ and $w_1=4$ are found by optimizing the categorical bias of model performance. 

We train our $0.65$M parameter model on a cluster of $8\times$ $40$ GB NVIDIA A100 GPUs for a maximum of $300$ epochs with AdamW. End-to-end training takes about $100$ hours wall-clock time, and due to high compute and limited accessibility, we perform minimal hyperparameter tuning (see \cref{tab:hyper} for final values) and report results of a single trial.

\section{Experimental Results} \label{sec:results}

Primary results and comparisons with the baseline model are discussed in \cref{sec:main}, and in \cref{sec:case} we investigate individual case studies of severe weather events (e.g., \cref{fig:cases}). Additional examples of model output in \cref{app:additional}.

\subsection{Main Findings} \label{sec:main}

We evaluate our results with three unique categories of methods, including: (a) \textit{standard metrics}: root-mean-squared error (RMSE) and coefficient of determination ($\text{R}^2$); (b) \textit{categorical metrics}: probability of detection (POD), false alarm ratio (FAR), and critical success index (CSI) at different composite reflectivity thresholds; and (c) \textit{sharpness quantification} with the mean magnitude of image gradients.

\paragraph{Model Performance} SRViT overall has better performance on the standard metrics, achieving an $\text{RMSE}=3.09\text{ dBZ and r}^2=0.572$, compared to the UNet's $\text{RMSE}=3.21\text{ dBZ and r}^2=0.488$ (\cref{tab:results}). These statistics capture general pixel-wise improvement, but are biased toward low coverage and zero-valued pixels. To improve our intuition of model performance, we evaluate the results on a linear interval where composite reflectivity is greater than predefined threshold levels. This more accurately displays where the improvements are captured.

Notably, in \cref{fig:stats}, we find SRViT exhibits the best performance at low- to mid-value thresholds. Relative to the UNet, the greatest improvements represented by FAR, CSI, and RMSE are between $[5, 40)$ dBZ. Specifically, there is a reduction of $0.96$ dBZ in RMSE for reflectivity greater than $35$ dBZ, while FAR increases by $0.08$ at a threshold of $25$ dBZ. However, the POD of each network is consistent across thresholds until a divergence occurs at $40$ dBZ. At these higher thresholds ($> 40$ dBZ), the UNet displays marginally better performance across all the categorical indicators, which may be attributed to specific test cases. 

\paragraph{Sharpness Evaluation} Qualitative sharpness can be defined by distinct, and sharp edge boundaries of neighboring pixels. An observation we find the lack of with the boundaries of composite reflectivity from the UNet. To quantify this observation, we compute the mean of the gradient magnitude for each output and compare the sample distribution among models. Specifically, we take the spatial mean after convolving a Sobel filter, i.e, \smash{$g=\frac{1}{m}\sum_{i=1}^m(G_{x_i}^2 + G_{y_i}^2)^{\frac{1}{2}}$}, where $x$ and $y$ are image directions, for each sample. \cref{fig:grad} shows the kernel density estimation (KDE) plots of $g$ across all test samples for each model. 

Both networks exhibit positively skewed distributions with lower mean values compared to MRMS. A direct comparison using Welch's independent samples t-test, show that SRViT demonstrates a statistically significant improvement in sharpness over the UNet ($\text{t-statistic}=27.71\text{, p-value} < 0.001$). These finding coincide with the qualitative assessments of observations (e.g., \cref{fig:cases}) and suggest that our method enhances the sharpness of composite reflectivity. However, both approaches still fall short of fully capturing the sharpness within the ground truth MRMS data. 

\subsection{Case Studies} \label{sec:case}

By enlarging model output, we study the results of our approach over a small spatial region for two distinct severe weather events and compare the predictions with MRMS and the UNet. In \cref{fig:cases}, the top row shows a Northern Plains Derecho and the bottom row shows Midwest Squall Lines. In both cases, SRViT has an improvement to the standard metrics (lower RMSE, higher $\text{R}^2$) relative to the UNet. A qualitative assessment has composite reflectivity boundaries with precise transitions, validating our findings of generating sharper output. The UNet also appears to overestimate for higher thresholds, whereas SRViT more accurately captures low- to mid-level thresholds but underestimates composite reflectivity $> 50$ dBZ, e.g., within the lower squall line over Southern Ohio in the bottom row of \cref{fig:cases}. These observations are consistent with the quantitative metrics we report in \cref{sec:main}.

\section{Interpreting Model Attribution} \label{sec:redistribution}

Some prior work aims to interpret the attention matrix of transformer models as importance or relevancy scores \cite{kovaleva2019revealing,clark2019does,carion2020end}. Indeed, these weights capture the interactions between all tokens, but faithful interpretations are challenged by not considering (a) the token vector magnitudes and high dimensionality and (b) the interplay between tokens across intermediate hidden layers \cite{brunner2019identifiability,wiegreffe2019attention,serrano2019attention}.

We address these issues by introducing Token (Re)Distri\-bution: an attribution method that captures the interaction of all other tokens, as learnt by the network, for a particular token. This approach allows for interpretations of how the value of input tokens are redistributed amongst the others. Based on prior work \cite{hao2021self,dhamdhere2018important,chefer2021transformer} that use the gradient of intermediate layers and self-attention features, we summarize gradient information across token embeddings to highlight the relevance of individual tokens (see details in \cref{app:trd}).

\begin{figure}[t]
    \centering
    \includegraphics[width=0.9\columnwidth]{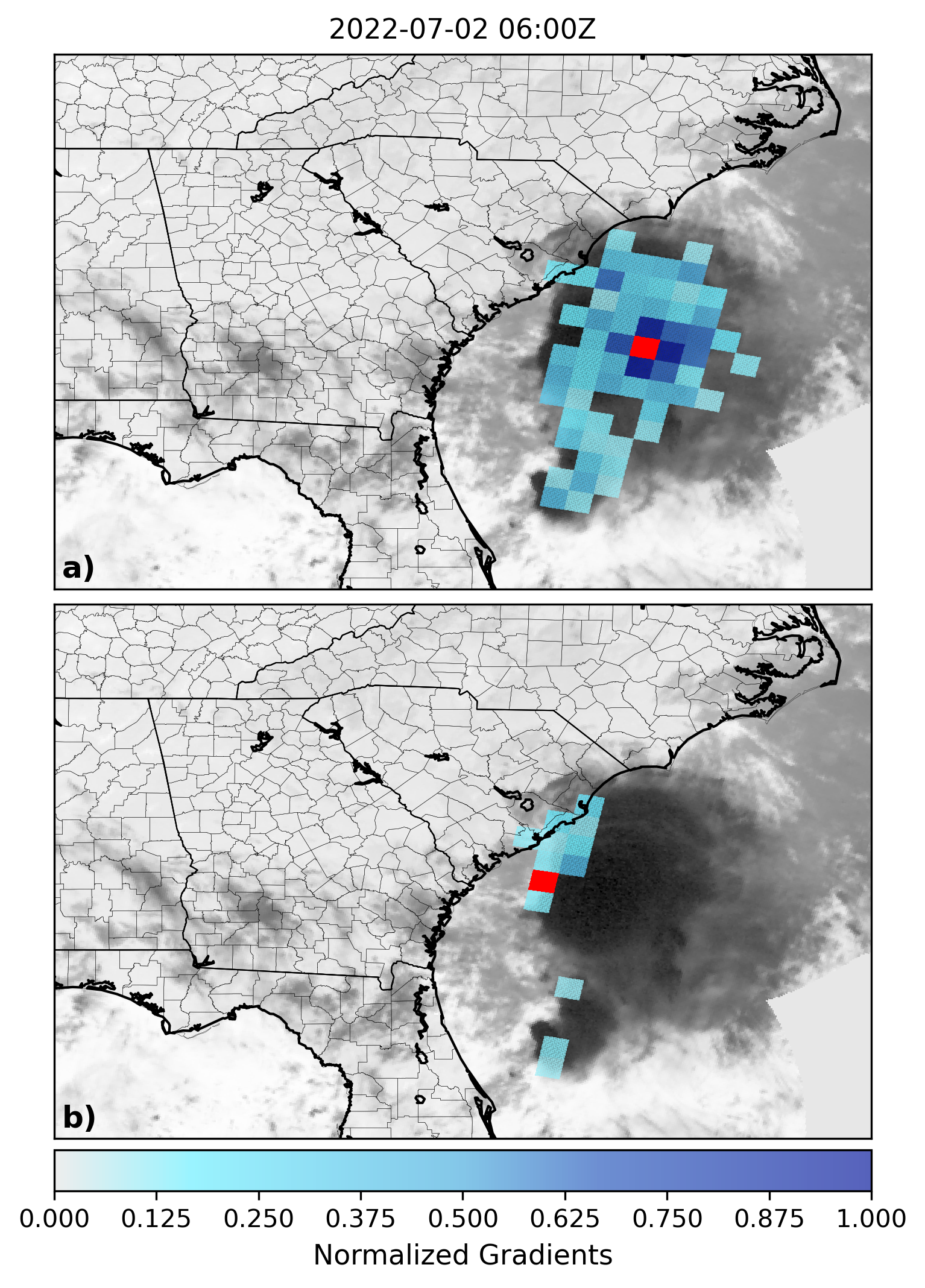}
    \caption{Sample min-max normalized gradient magnitude of Token (Re)Distribution for the (red; separate in panels (a) and (b)) token of interest overlaid on GOES-16 ABI Channel 7.}
    \label{fig:redistribution}
\end{figure}

An example attribution map with model output of Tropic Storm Colin is shown in \cref{fig:redistribution}. Each panel shows the result of a different token, from the same sample, with the normalized gradient magnitude. In \cref{fig:redistribution}a the token of interest, centered on higher radiance values, shows surrounding tokens within the clouds contributing to its value. Similarly, in \cref{fig:redistribution}b, the token on the storm edge uses the values of others along this edge. Both cases are meteorological supported as it relates to the storm structure and extent of precipitation. Altogether, this method is a step toward guiding domain experts' interpretations of model output.

\section{Conclusion} \label{sec:conclusion}

Our analyses demonstrate that SRViT improves low- and mid-value estimates of composite reflectivity and produces sharper predictions compared to convolutional approaches. Case studies validate the model's capability to accurately capture complex atmospheric phenomena both locally and across larger domains. Additionally, our gradient-based attribution method provides valuable meteorological insights and is applicable to various transformer architectures.

Despite these advancements, some limitations remain. The UNet outperforms SRViT at high-value estimates of composite reflectivity, and neither model fully achieves the sharpness of MRMS. Future work could address these gaps by exploring probabilistic diffusion models to better capture uncertainty and sharpness while also developing methods to estimate 3D radar reflectivity fields to better inform NWP.



\textit{Code available on GitHub}: \href{https://github.com/stockeh/srvit}{github.com/stockeh/srvit}

\section*{Acknowledgments}
This work is supported by NSF Grant No. 2019758, \textit{AI Institute for Research on Trustworthy AI in Weather, Climate, and Coastal Oceanography (AI2ES)} as well as the GOES-R Program under grant NA19OAR4320073. A special thanks to the University at Albany for allowing us to use their xCITE A100 GPU compute cluster for experiments and model training. 

\bibliography{example_paper}
\bibliographystyle{icml2024}

\newpage
\appendix
\onecolumn

\section{Dataset Details} \label{app:dataset}

\cref{app:dataset.varaibles} overviews the critical data variables and properties unique to this study; however, we refer the reader to \citet{Hilburn2021Gr} for more complete data specifications. We also provide insights to the spatial coverage in \cref{app:dataset.coverage}.

\subsection{Variable Description} \label{app:dataset.varaibles}

\paragraph{Advanced Baseline Imager (ABI)} We use Level-L1b radiances from GOES-16 ABI \cite{schmit2010goes}. The imagery has a nominal spatial resolution of $2$ km, enabling the capture of fine-grained details, while ensuring a temporal refresh rate of $5$ minutes for timely data updates. To account for both day-night scenarios, we concentrate on the use of the infrared bands, specifically: \textbf{Channel 7} (\SI{3.7}{\micro\metre}; Shortwave Window), \textbf{Channel 9} (\SI{6.9}{\micro\metre}; Upper-Level Water Vapor), and \textbf{Channel 13} (\SI{10.3}{\micro\metre}; Clean IR Longwave Window).

\paragraph{Geostationary Lightning Mapper (GLM)} Accompanying the satellite imagery are real-time lightning observations from the GOES-R GLM \cite{goodman2010glm,goodman2013goes}. This instrument is a single-channel, near-infrared transient detector that monitors total lightning with a uniform spatial resolution of approximately $10$ km. Lightning itself is particularly useful for generating synthetic radar fields as their locations are often associated with strong updrafts within convective environments.

\paragraph{Multi-Radar Multi-Sensor (MRMS)} We utilize quality-controlled composite reflectivity from the MRMS product \cite{smith2016multi} as our target dataset. This data combines information from different radar networks, surface observations, numerical weather prediction (NWP) models, and climatology. Through a sophisticated integration process, high-resolution mosaics are generated, offering precise spatiotemporal resolution. However, it is important to note that the coverage of the product is both limited in the vertical and spatial domain due to beam blockage and radar placement.

\subsection{Data Coverage and Distribution} \label{app:dataset.coverage}

\begin{figure*}[ht!]
    \centering
    \begin{minipage}[b]{0.475\textwidth}
        \centering
        \includegraphics[width=0.9\textwidth]{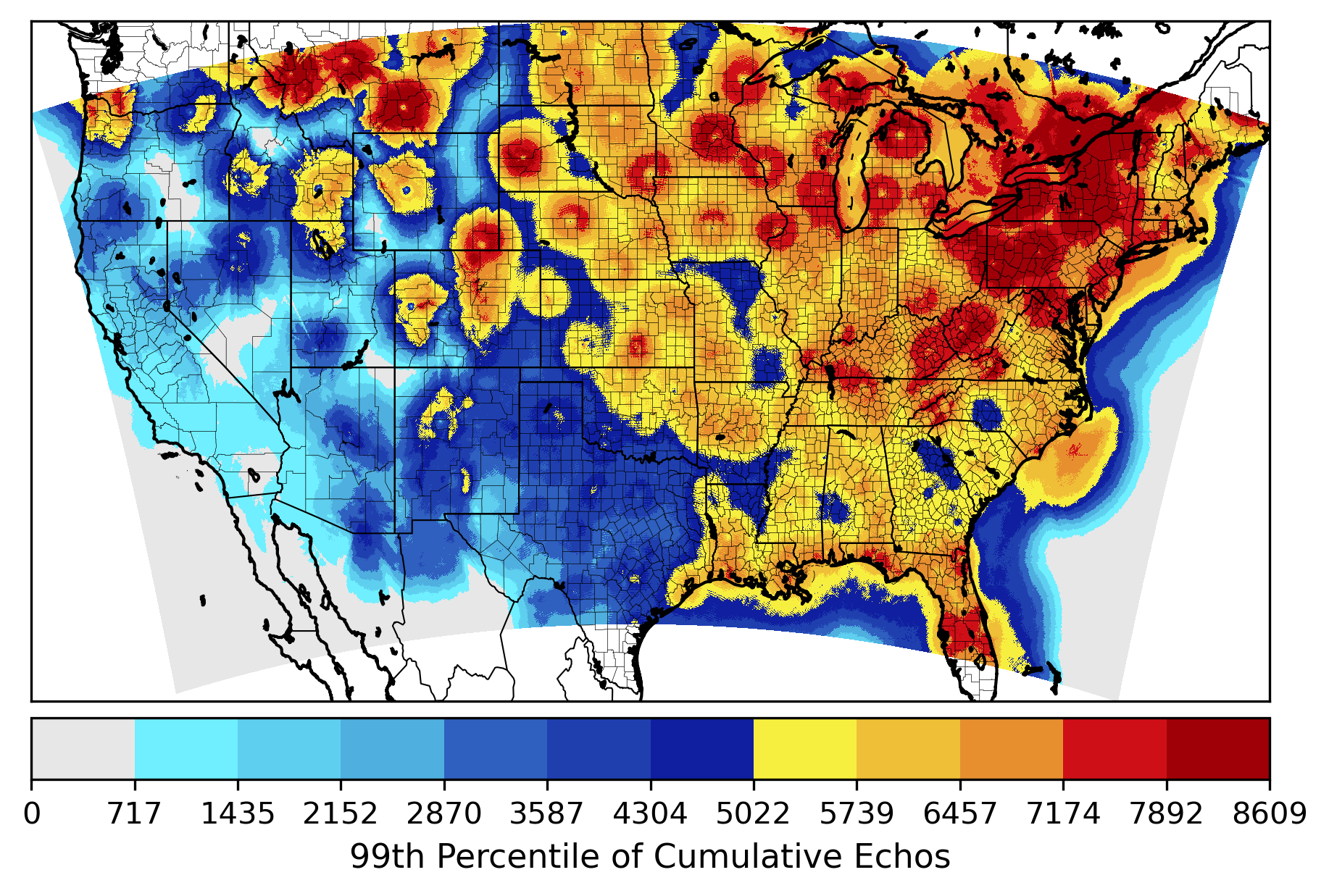}
        \caption{Spatial coverage of MRMS from the training data.}
        \label{fig:spatial-distribution}
    \end{minipage}
    \hfill
    \begin{minipage}[b]{0.475\textwidth}
        \centering
        \includegraphics[width=0.9\textwidth]{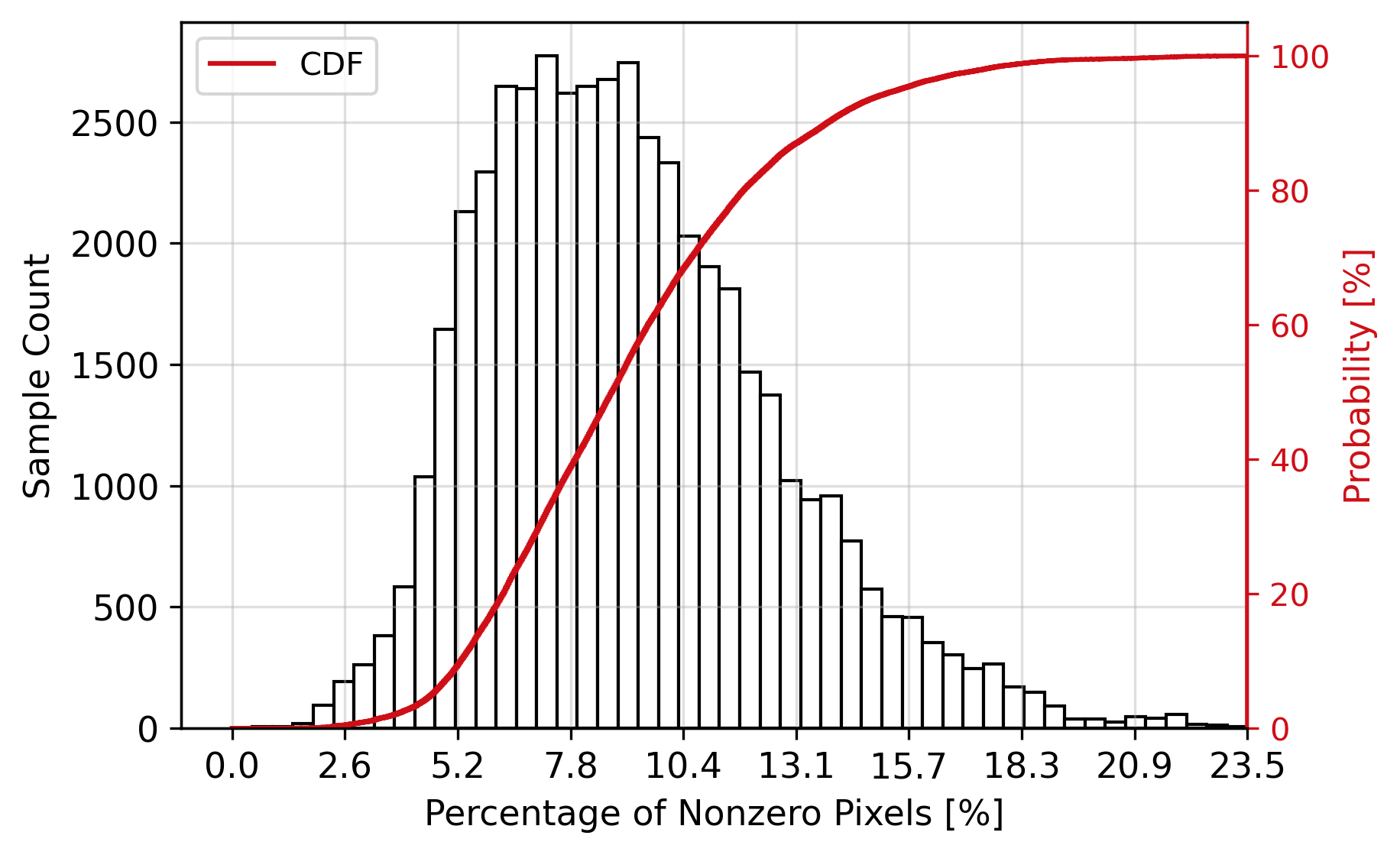}
        \caption{Sample histogram and cumulative density function of nonzero pixels present in the training data.}
        \label{fig:cdf}
    \end{minipage}
\end{figure*}

\cref{fig:spatial-distribution} illustrates the radar coverage from the training data, presenting a spatial map of cumulative echos. To represent the spatial distribution, we binarize the data by assigning a value of $1$ to composite reflectivity values greater than $0$, which are then summed across all samples. We then clip the result based on the $99{\text{-th}}$ percentile to eliminate visual artifacts. Notably, the coverage is lowest over the western United States, with a bias towards physical radar stations. The western region, affected by orographic enhancement, is further impeded by mountains (i.e., subject to beam blockage), resulting in predominantly high-level atmospheric echoes that largely miss the stronger echoes near the surface. In the most extreme cases, precipitation may go completely undetected by radars.

In order to disentangle the sample specific coverage from the cumulative spatial representation, we compute the sample frequency of nonzero pixels with its cumulative density function (CDF). \cref{fig:cdf} displays the result on the training data, showing a histogram with a slight positive skew and mean centered around $9.2\%$, indicating moderate coverage within samples relative to the CONUS domain. Moreover, a substantial proportion of the training samples have nonzero pixel percentages above the median value. Importantly, this indicates that all samples have some level of coverage, irrespective of their spatial extent, without exceeding into high, erroneous values.

\begin{table*}[b!]
    \centering
    \begin{minipage}[b]{0.475\textwidth}
        \centering
        \begin{small}
            \begin{sc}
                \begin{tabular}{lc}
                \toprule
                Hyperparameter & Value \\
                \midrule
                Global Batch Size & 16 \\
                Learning Rate $\eta$ & $1e-4$ \\
                Epochs (patience) & 300 (50) \\
                \midrule
                Patch Size $(p\times p)$ & $(12 \times 12)$ \\
                Model Depth $L$ & 6 \\
                Heads (per block) & 12 \\
                Model Dimension $d$ & 256 \\
                Inner Dimension $v$ & 64 \\
                FCN Dimension $m$ & 512 \\
                Conv Hiddens (filters) & $[32, 16]$\\
                \bottomrule
                \end{tabular}
            \end{sc}
        \end{small}
        \caption{SRViT hyperparameters ($643{,}105$ parameters).}
        \label{tab:hyper}
    \end{minipage}
    \hfill
    \begin{minipage}[b]{0.475\textwidth}
        \centering
        \begin{small}
            \begin{sc}
                \begin{tabular}{lccc}
                \toprule
                Model & $\downarrow$ RMSE (dBZ) & $\uparrow \text{R}^2$ & $\uparrow$ Sharpness ($g$)\\
                \midrule
                MRMS & -- & -- & 0.48 $\pm$ 0.16 \\
                UNet & 3.21 & 0.488 & 0.21 $\pm$ 0.09 \\
                Base-ViT & 3.05 & 0.487 & 0.21 $\pm$ 0.09 \\
                \midrule
                SRViT & 3.09 & 0.572 & 0.24 $\pm$ 0.11 \\
                \bottomrule
                \end{tabular}
            \end{sc}
        \end{small}
        \caption{Standard and sharpness metrics across models.}
        \label{tab:results}
    \end{minipage}
\end{table*}

\section{Architectural Details} \label{app:model}

Consider an image $I \in \mathbb{R}^{c \times h \times w}$ that is partitioned into equally divisible patches (synonymous with tokens) of size $p$ as $x = \{x_0, \dots, x_n\}$, where $x_i \in \mathbb{R}^{d_{in}}$ and $n = (w / p)\cdot(h / p)$ with $d_{in} = p^2 c$. Once partitioned, the set of tokens $x$ are linearly projected to dimension $d < d_{in}$ and we add a standard sine-cosine positional encoding. Let matrix $\mathbf{X}^l \in \mathbb{R}^{n \times d}$ be the new row-wise concatenation of the tokens. A typical transformer block $\phi$ at layer $l$ processes the set of tokens with multi-head self attention (MSA) and a point-wise fully-connected network (FCN) as,
\begin{align} 
    \phi(\mathbf{X}) &= \text{FCN}(\text{MSA}(\mathbf{X})) \quad \text{such that} \label{eq:block} \\
    \text{MSA}(\mathbf{X}) &= [\mathbf{O}_1, \mathbf{O}_2, \dots, \mathbf{O}_h]\mathbf{W^O} \label{eq:msa},
\end{align}
where $h$ is the number of heads, $\mathbf{W^O} \in \mathbb{R}^{hv \times d}$ are trainable weights, $[\cdot]$ is the column-wise concatenation, and $\mathbf{O}_i \in \mathbb{R}^{n \times v}$ is the output of the $i$-th attention head with latent dimension $v < d$. We compute each head as,
\begin{align} 
    \mathbf{O}_i &= \mathbf{A}_i\mathbf{V}_i \quad \text{such that} \label{eq:output} \\
    \mathbf{A}_i &= \text{softmax}\big(\mathbf{Q}_i\mathbf{K}_i^\top / \sqrt{d}\big) \in \mathbb{R}^{n \times n}. \label{eq:attn}
\end{align}

The queries, $\mathbf{Q}_i$, keys, $\mathbf{K}_i$, and values, $\mathbf{V}_i$ are found via a linear projection of $\mathbf{X}$ by,
\begin{equation} \label{eq:qkv}
    \mathbf{Q}_i = \mathbf{X}\mathbf{W}^Q_i, \quad \mathbf{K}_i = \mathbf{X}\mathbf{W}^K_i, \quad \mathbf{V}_i = \mathbf{X}\mathbf{W}^V_i,
\end{equation}
with trainable weight matrices $\mathbf{W}^Q_i, \mathbf{W}^K_i, \mathbf{W}^V_i \in \mathbb{R}^{d \times v}$. The FCN block at layer $l$ is a two layer network separated by the Gaussian Error Linear Unit (GELU) activation, $\delta$, and dropout (with a default $p=0.2$) that takes as input the layer normalized output of the MSA, $\bar{\mathbf{X}} = \text{LN}(\mathbf{X})$. This block is computed as,
\begin{equation} \label{eq:fcn}
    \text{FCN}(\bar{\mathbf{X}}) = \delta(\bar{\mathbf{X}}\mathbf{W}^R)\mathbf{W}^S,
\end{equation}
where $\mathbf{W}^R \in \mathbb{R}^{d \times m}$ and $\mathbf{W}^S \in \mathbb{R}^{m \times d}$ such that $m > d$.

Each transformer block has independent weight matrices and yields a new set of tokens of the same dimension given by $\phi : \mathbf{X}^l \rightarrow \mathbf{X}^{l+1}$. After $L$ transformations, we introduce a linear decoding block and reshaping operation $f$ to reconstruct the intermediate output. This is denoted by,
\begin{equation} \label{eq:head}
    z_0 = f(\mathbf{X}^L) = \text{reshape}(\mathbf{X}^L\mathbf{W}^F , (c \times h \times w)),
\end{equation}
where $\mathbf{W}^F \in \mathbb{R}^{d \times d_{in}}$ and $z_0$ is the same dimension as the input image $I$. The output of this layer is a close approximation to the output, but as evidence in \cref{sec:ablation}, we show it to be inaccurate. Therefore, we include convolutional layers following $f$ to effectively smooth the boundaries of the decoded tokens via weight sharing. We represent the $N$ subsequent convolutional layers as,
\begin{equation}
    z_i = \text{ReLU}(\mathbf{W}^C_i * z_{i-1}), \text{ for } i = 1 \text{ to } N,
\end{equation}
where $\ast$ denotes the convolution operation, $\mathbf{W}^C_i$ are the trainable weights at layer $i$, and ReLU is applied element-wise. The model's final output, $y$, when $i = N$ has the same spatial dimension as $I$ with a single output channel.

\newpage
\section{Additional Model Output} \label{app:additional}

\begin{figure*}[!ht]
    \centering
    \includegraphics[width=1\textwidth]{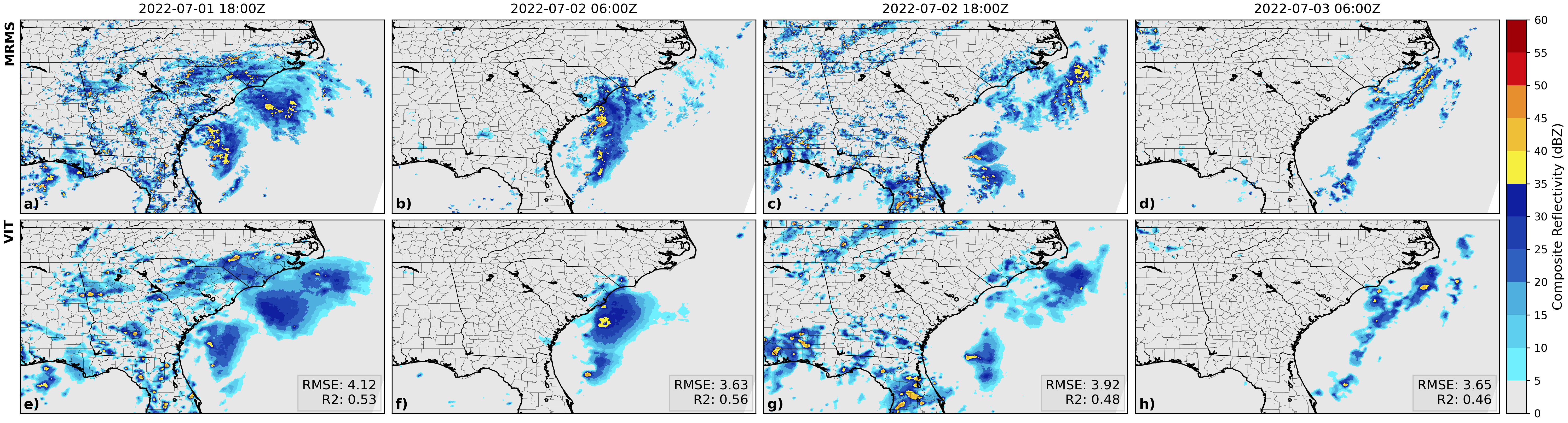} 
    \vspace{-4mm}
    \caption{Example model output, showing the temporal progression of Tropical Storm Colin as it compares MRMS (top row) and SRViT (bottom row). Sample RMSE and $\text{R}^2$ values are shown for each timestep (computed between individual panel columns).}
    \label{fig:tc-colin}
\end{figure*}

\cref{fig:tc-colin} illustrates the progression of Tropical Storm Colin on 12 hour intervals from the output of SRViT on an Equidistant Cylindrical map projection. The short-lived storm had a peak intensity of $35$ knots from 2330 UTC July 1, 2023 to 1200 UTC July 2, 2023 over the western Atlantic. Accompanied with deep convection, Colin brought at most $19.28$ cm of rainfall to Wadmalaw Island, South Carolina \cite{latto2022colin}. SRViT adequately captures Colin's structure (\cref{fig:tc-colin}e) while accurately representing inland scattered convection (\cref{fig:tc-colin}g,h). This observation demonstrates its ability to handle complex weather patterns over the entire CONUS domain.

\section{Ablation Study} \label{sec:ablation}

To better understand our network design, we ablate the key components and discuss the qualitative and quantitative implications. Following the transformer blocks in SRViT are convolutional layers that effectively smooth the boundaries of decoded tokens. By training the base transformer model with the same hyperparameters and training procedure but without convolutions, we can study this observation. 

We find quantitative statistics of the base model having an $\text{RMSE}=3.05\text{ dBZ and r}^2=0.487$ (see \cref{tab:results}). This shows a lower pixel-wise error, which means its predictions are, on average, closer to MRMS. However, the lower $\text{R}^2$ suggests that the use of convolutions better explains a larger proportion of the variance and captures the underlying patterns more effectively. When assessing sharpness, it becomes evident that the base model, having a mean gradient magnitude of $0.21 \pm 0.09$, fails to capture the subtle transitions of composite reflectivity boundaries.

\begin{figure}[!ht]
    \centering
    \includegraphics[width=.85\linewidth]{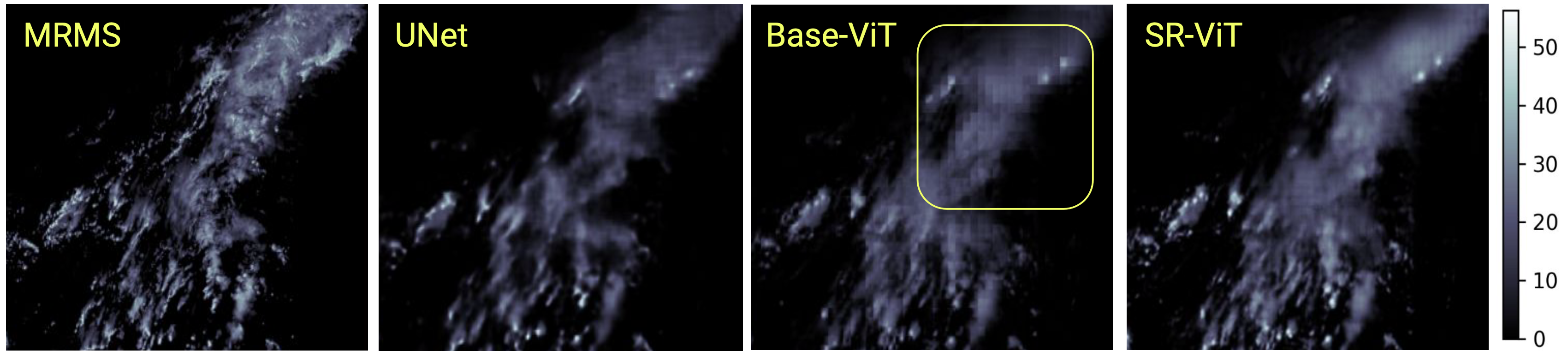}
    \caption{Observational ``patchiness'' most evident within the yellow box across model architectures.}
    \label{fig:patchiness}
\end{figure}

Through qualitative observations (\cref{fig:patchiness}), it is evident that the base model produces output with noticeable tiling or ``patchiness''. We suspect this artifact arises from the process of decoding each token individually, which negatively impacts the natural transition and coherence of the output. By contrast, incorporating additional convolutional layers proves to be effective in capturing intricate details and avoiding the tiling issues, making it a better choice for generating synthetic radar.

\section{Details of Token (Re)Distribution} \label{app:trd}

The activation of scaled dot-product attention computes the matrix $\mathbf{A} \in \mathbb{R}^{n \times n}$ that represents the reweighing of our values from the combined queries and keys (see \cref{eq:output,eq:attn}). Recall that each transformer block yields an activation matrix as given by $\phi : \mathbf{X}^l \rightarrow \mathbf{X}^{l+1} \in \mathbb{R}^{n \times d}$. Let the input token embedding prior to adding the positional encoding be $\mathbf{X}$ and the output of the first block be $\mathbf{Z}$ (this can be for any block following the rules for backpropagation). The output matrix $\mathbf{U} \in \mathbb{R}^{n \times n}$ has for the $i$-th intermediate token (as a row-vector, $\mathbf{z}$), the sensitivity to all input tokens in the $j$-th column. By differentiating $\partial \mathbf{z} / \partial \mathbf{X}$, we encode the partial derivatives in a third-order tensor, $\mathbf{D}^\mathbf{z} \in \mathbb{R}^{d \times n \times d}$, indexed by $\mathbf{D}^\mathbf{z}_{i,j,k} = \partial \mathbf{z}_i / \partial \mathbf{X}_{jk}$. In practice, we compute this efficiently as the vector-Jacobian product with the row-vector $\mathbf{1} = [1_1,\dots,1_d]$ that reduces the tensor to $\mathbf{D}^\mathbf{z} \in \mathbb{R}^{n \times d}$. We summarize the computation with all $n$ row-vectors in $\mathbf{Z}$ as,
\begin{align} 
    \mathbf{U} &=  [f(\mathbf{Z}_1)_1,\dots,f(\mathbf{Z}_n)_n] \quad \text{such that} \label{eq:routput} \\
    f(\mathbf{z})_i &= \Big\lvert\sum\nolimits_{j=1}^d \mathbf{D}^\mathbf{z}_{ij}\Big\rvert \in \mathbb{R}^{n}. \label{eq:rsum}
\end{align}

The function $f$ reduces the $d$-dimensional gradients of each token's sensitivity for all $n$ input tokens. This provides a  holistic view for how the change in the input token, over its entire embedding, affects the intermediate token; a large magnitude is highly responsive to small changes. Thus, we define a redistribution of those input tokens, as a result of self-attention, to the value of an intermediate token. With multiple transformer blocks, we can take the mean over all blocks for a more complete view. Thereafter, visualization of this mean are made using data indexed at a given row in $\mathbf{U}$.

\end{document}